\documentstyle[aps]{revtex}
\begin{document}
\voffset=0mm
%
\title{THE TWO - DIMENSIONAL ATTRACTIVE HUBBARD MODEL: HIGHLY NON-LINEAR 
SUPERCONDUCTIVITY WITH SUM RULES.}
\author{J.J. Rodr\'{\i}guez-N\'u\~nez} 
\address{Departamento de F\'{i}sica, Universidade Federal de Santa Maria, 
CCNE, 97105 - 900 Santa Maria/RS, Brazil.\\
e-m: jjrn@mfis.ccne.ufsm.br}
\author{C.E. Cordeiro and A. Delfino}
\address{Instituto de F\'{\i}sica, Universidade Federal Fluminense, 
Av.\ Litor\^anea S/N, Boa Viagem, 24210-340 Niter\'oi/RJ, Brazil.}
\date{\today}
\maketitle

%
%
\begin{abstract}
We use the moment approach of Nolting (exact sum rules) 
(Z. Physik {\bf 255}, 25 (1972)) for the attractive 
Hubbard model in the superconducting phase. Our diagonal 
and off - diagonal spectral functions are constructed and 
evaluated with the sum rules. They reduce to the $BCS$ limit 
for weak interaction. However, the presence of correlations 
modify the $BCS$ picture dramatically. For example, due to 
the presence of correlations we have postulated a three - 
pole ansatz for the diagonal Green function, $G(\vec{k},\omega)$, 
while the off - diagonal one, $B(\vec{k},\omega)$, 
is supposed to have two poles. In the paper we present results 
for the three spectral weights of the diagonal Green 
function, $\alpha_j(\vec{k})$, j = 1,2,3. 
Our results compare reasonably well with more 
elaborated auto - consistent highly non - linear equations 
(double fluctuation calculations in the $T-$ Matrix approach of 
one of the authors). Then, the physical picture which emerges 
is that the lower Hubbard band is split due to the superconducting 
gap and the upper Hubbard band remains mostly unmodified.\\
\\
Pacs numbers: 74.20.-Fg, 74.10.-z, 74.60.-w, 74.72.-h
\end{abstract}

\pacs{PACS numbers 74.20.-Fg, 74.10.-z, 74.60.-w, 74.72.-h}
%
%
%
%
\section{Introduction}\label{sec:intro}

	The study of correlations has been renewed again by the
the discovery of high $T_c$ superconducting oxide
materials (HTSC),\cite{Bednorz-Muller} since
that these materials exhibit a short coherence
lenght, $\xi$, and a very large penetration depth, $\lambda$,
which become them almost extreme type-II superconductors, i.e.
$\kappa \equiv {{\lambda} \over {\xi}} \gg 1$.

    	In order to study the effect of
correlations we will use the on-site attractive Hubbard Hamiltonian
put forward by Micnas et al \cite{Micnas_et_al} as a
phenomenological model for describing the HTSC. Previous authors 
have used this model to study the bismuthate superconductors.\cite{3} 
Denteneer et al \cite{Denteneer} recognize
that if the phase diagram of the Hubbard model were fully
understood, it might form the basis of understanding correlated
electrons as much as the Ising model did for understanding
critical phenomena. More recently,
Schneider et al\cite{Schneider_et_al} have applied
this model to explain universal properties of several
families of these compounds, like the relation between transition
temperature, magnetic penetration depth and gap, at zero
temperature.

	In this paper, we will use the exact relations of
Nolting\cite{Nolting} for the one-particle diagonal spectral
function together with the exact relations for the one-particle
off-diagonal spectral function, i.e., the anomalous Green's
function, to study systems with broken symmetry.\cite{KF}

    	The model we study is the following
\begin{eqnarray}
\label{Ham}
H = - t\sum_{<i,j>\sigma}c_{i\sigma}^{\dagger}c_{j\sigma}
   + U \sum_i n_{i\uparrow}n_{i\downarrow}
   - \mu \sum_{i\sigma} n_{i \sigma}~~,
\end{eqnarray}
where $c_{i\sigma}^{\dagger}$($c_{i\sigma}$) are creation
(annihilation) electron operators with spin $\sigma$. $n_{i
\sigma} \equiv c_{i\sigma}^{\dagger}c_{i\sigma}$, $t$ is a
hopping matrix element between the nearest sites (n.n.) $i$
and $j$, and $U$ is the onsite interaction. $\mu$ is the chemical
potential and we work in the grand canonical ensemble. In the
present study we consider an attractive interaction,
$U \equiv - \mid U \mid ~<~ 0$. The Hamiltonian
of Eq.~(\ref{Ham})~has been studied in
detail by Micnas et al\cite{Micnas_et_al}. Nolting\cite{Nolting} 
and Kalashnikov and Fradkin\cite{KF} have derived exact relations 
for the one-particle spectral functions which are given in 
Ref.\cite{PhysicaA1}.

	We postulate for the diagonal spectral function, 
$A(\vec{k},\omega)$, the following ansatz
\begin{equation}\label{diaspec}
A({\bf k},\omega) \equiv \alpha_1(\vec{k})\delta(\omega - 
E_{\vec{k}}) +  \alpha_2(\vec{k})\delta(\omega +  
E_{\vec{k}}) +  \alpha_3(\vec{k})\delta(\omega - 
\Omega_{\vec{k}})~~~
\end{equation}

\noindent
and the off-diagonal spectral function, $B(\vec{k},\omega)$, is 
supposed to have two peaks ($BCS$-like), as follows

\begin{equation}\label{offspec}
B(\vec{k},\omega) \equiv \sqrt{\alpha_1(\vec{k}) \alpha_2(\vec{k})} 
\left[\delta(\omega - E_{\vec{k}}) - \delta(\omega + E_{\vec{k}})
\right] ~~~,~~~\Omega_1(\vec{k}) = E_{\vec{k}}~~~,~~~
\Omega_2(\vec{k}) = - E_{\vec{k}}~~~
\end{equation}

	Our ansatz is based on the assumption that the 
diagonal spectral function is modified by the presence of 
correlations (we define correlations when we have the 
regime $U/W \approx 1$, where $W = 8t$  is the bandwidth in 
two dimensions). This assumption has been numerically 
checked to be correct in Ref.\cite{Dflu} where the authors 
have obtained three bands for the diagonal spectral function 
and two bands for the off-diagonal spectral function after 
performing the analytical continuation. \noindent
In Eq. (\ref{diaspec}), the first two poles represent 
the behavior around the chemical potential and the
third pole is due to the influence 
of the upper Hubbard band, which in the 
case of the atomic limit is the upper Hubbard band or the 
band of single occupied states. The lower
band, now split in two, is the band of 
double occupied states for the case of allmost 
atomic limit. Please, see reference \cite{Micnas-et-al}, where 
the attractive Hubbard model has been studied in the normal 
phase with two different methods: $T$-Matrix and moments.

	The order parameter, $\Delta_o(T)$, 
and the chemical potential (fixed electron density), 
$\mu$, are calculated from the
following equations:\\
\begin{equation}\label{2selfeq}
\Delta_o(T) = \frac{1}{N} \sum_{\vec{k}}\int_{-\infty}^{+\infty} 
\frac{B(\vec{k},\omega)}
{e^{\beta \omega}+1}~~~; ~~~ \rho =  \frac{1}{N} 
\sum_{\vec{k}}\int_{-\infty}^{+\infty} 
\frac{A(\vec{k},\omega)}{e^{\beta \omega}+1} ~~~.
\end{equation}
	For example, the quasi-par\-ti\-cle energy, 
$E_{\vec{k}}$, has a gap. As it was discussed in 
Ref.\cite{PhysicaA1}, the gap is expressed in 
terms of the order parameter as
\begin{equation}\label{k-gap}
\Delta(T,\vec{k}) = \Delta_0(T) \frac{2E_{\vec{k}}}
{E_{\vec{k}}+a_1(\vec{k})}~~~,
\end{equation} 
\noindent
concluding from Eq. (\ref{k-gap}) that our gap equation, 
which is a manifestation of singularities in the density of states, 
is $\vec{k}$-dependent. Going back to our local Hamiltonian 
(see Eq. (\ref{Ham})), which in reciprocal space is a constant, then 
at the mean field level we should obtain a {\it pure s-wawe}, while 
going beyond mean field approximation we have been able to derive 
a $\vec{k}$-dependent gap. In our case, by including correlations, 
i.e., the selection rules for $a_2$ and $a_3$ (See 
Ref.\cite{PhysicaA1} for more details) we have been able to 
modify the BCS results, from pure s-wave to 
a wave vector dependent energy gap. Let us point out that 
in the attractive Hubbard model we always find $s-$type 
wave symmetry as it has conclusively shown in Ref.\cite{india} 
in the Hubbard-I approximation for both the diagonal and the 
off-dia\-go\-nal one-par\-ti\-cle Green functions.

	We would like to call the reader's attention that our Ansatz is 
based on the assumption that the role of correlations is mainly 
taken into account in the diagonal one-particle spectral function. This 
implies that the off-diagonal order parameter has been taken at the  
BCS level, i.e., to $M^{(1)}({\bf k})$. In a previous 
work,\cite{all} it is found that both the diagonal and 
off-diagonal spectral functions have four peaks, symmetric in 
pairs, for 
$U/t = -4$. However, when we have included double fluctuations, i.e., 
fluctuations in both the diagonal and off-dia\-go\-nal 
self-e\-ner\-gies as in Ref.\cite{Dflu}, we have three resolved peaks  
for $A(\vec{k},\omega)$ and two well resolved peaks for 
$B(\vec{k},\omega)$. So, Ref.\cite{Dflu} has motived us to adopt 
three and two peaks for $A(\vec{k},\omega)$ and $B(\vec{k},\omega)$, 
respectively, for $U/t = - 4.0$.  We have restricted ourselves to 
wavevectors close to the chemical potential. If we want to 
consider two other peaks for the off-diagonal spectral function we 
have to include more moments, which is beyond the present work. 
Also, we have neglected life-time effects in Eqs. 
(\ref{diaspec}-\ref{offspec}), 
which would require to approach the phase transition with damping 
in the self-e\-ner\-gy as it has been done in Ref.\cite{MMIT}.

	The equations we have to solve are the following

\begin{eqnarray}\label{self1}
\frac{1}{U} &=& - \frac{1}{N} \sum_{\bf k} \frac{
\tanh(\frac{\beta E_{\bf k})}{2})}{2E_{\bf k}} \\
\rho &=& ~\frac{1}{N} \sum_{\bf k} \left[ \alpha_1({\bf k}) f(E_{\bf k})
+ \alpha_2({\bf k}) f(-E_{\bf k}) + \frac{a_1({\bf k}) - 
\Omega_1({\bf k})}{\Omega_2({\bf k}) - 
\Omega_1({\bf k})}f(\Omega_{\bf k}) \right] ~~~
\end{eqnarray}

In Fig. 1 we show the dependence of the diagonal spectral weigths, 
$\alpha_j(\vec{k})$, j=1,2,3 along the diagonal of the Brillouin 
zone. We can appreciate that the area of the spectral function 
is equal to one ($\alpha_1 + \alpha_2 + \alpha_3 = 1$) by 
construction. We have taken $U/t = - 4.0$, $\rho = 0.1$ and 
$T/t = 0.01$. After self-con\-sis\-ten\-cy of our highly 
non-\-li\-near equations we obtain $\mu/t = - 4.19$, $B/t = 
1.664$ and $\Delta_0(T) = 0.52$. It is worth mention that 
$\mu < -4t$ which indicates that we are in the low density 
regime (Bose - Einstein limit). In Fig. 2 we have the 
diagonal spectral weights along the $k_x$-axis ($\vec{k} = 
(k,0)$). What we observe is a strong variation of the 
$\alpha_j(\vec{k})$ with direction. In Fig. 3 we have the 
$\vec{k}$-de\-pen\-den\-ce of $\gamma_j(\vec{k})$, 
j = 1,2 along the two directions as previously discussed in 
Figures 1 and 2. These $\gamma_j(\vec{k})$'s 
are defined by Eq. (19) in Ref.\cite{PhysicaA1}. They are 
highly $\vec{k}$-de\-pen\-dent. The meaning of these 
variables is that they represent corrections to the diagonal 
sum rules ($a_2(\vec{k})$ and $a_3(\vec{k})$). As we have 
said in Ref.\cite{PhysicaA1} these corrections are important, 
i.e., of the order of 20 - 25 
higher corrections on the order parameter (or equivalently, 
$\gamma_j({\vec{k}})$'s). For $T \approx T_c$ these corrections 
are not important, though.

	By using the moment approach  in the presence
of off-diagonal long range order, i.e.,$\Delta_o(T) \neq 0$, to the negative Hubbard model, we have worked out the superconducting phase calculated 
with the use of moments (exact sum rules for both the diagonal and 
off-dia\-go\-nal spectral densities) with three and two  
and two peaks, respectively. The physical meaning of each of these 
peaks has been discussed. This assumption is equivalent to solve the 
diagonal spectral weights in powers of the order parameter square. At 
the same time, we have seen that the effect of the third band is to 
renormalize the order parameter producing an energy gap which is 
$\vec{k}$-dependent. In consequence, in our approach the 
order parameter, $\Delta_o(T)$, and the energy gap, 
$\Delta(T,\vec{k})$ are not the same. We would like to end by 
saying that there are strictly mean - field calculations (even if 
they go beyond Hartree - Fock) as the ones done by Mart\'{\i}n - 
Rodero and Flores\cite{9} and Kuchiev and Sushkov\cite{10}. Our 
approach presented here is able to capture the double fluctuaction 
calculation of Ref.\cite{Dflu}. This clearly indicates that we are beyond 
a simple $BCS$ calculations.

    One of the authors (JJRN) would
like to acknowledge partial support from CONDES-LUZ
and also from CONICIT (project F-139). 
We thank Mar\'{\i}a Dolores Garc\'{\i}a
for a reading of the manuscript.\\

\newpage
\begin{center}
{\Large Figures}
\end{center}

\vspace{0.5cm}
\noindent Figure 1. The spectral weights, $\alpha_j(\vec{k})$, 
$j=1,2,3$ along the diagonal of the Brillouin zone. The first diagonal 
sum rule is satisfied. $U/t = -4.0$, $\rho = 0.1$ and $T/t = 0.01$.\\

\noindent Figure 2. The spectral weights, $\alpha_j(\vec{k})$, 
$j=1,2,3$ along the $k_x$ direction of the Brillouin zone. Same 
parameters as in Figure 1.\\

\noindent Figure 3. $\gamma_j(\vec{k})$, $j = 1,2,3$ along the two 
directions discussed in Figures 1 and 2. Same parameters as before.\\
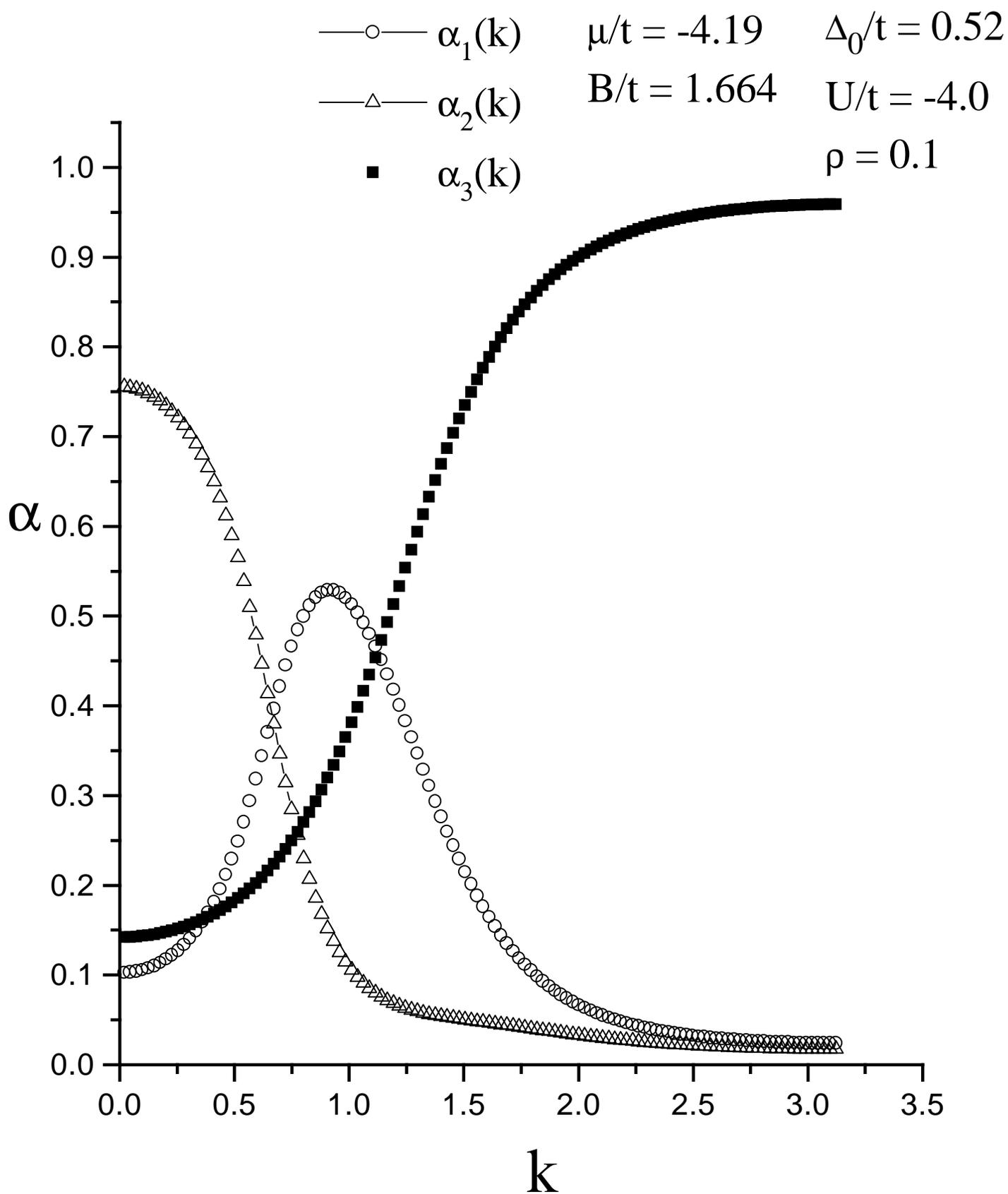
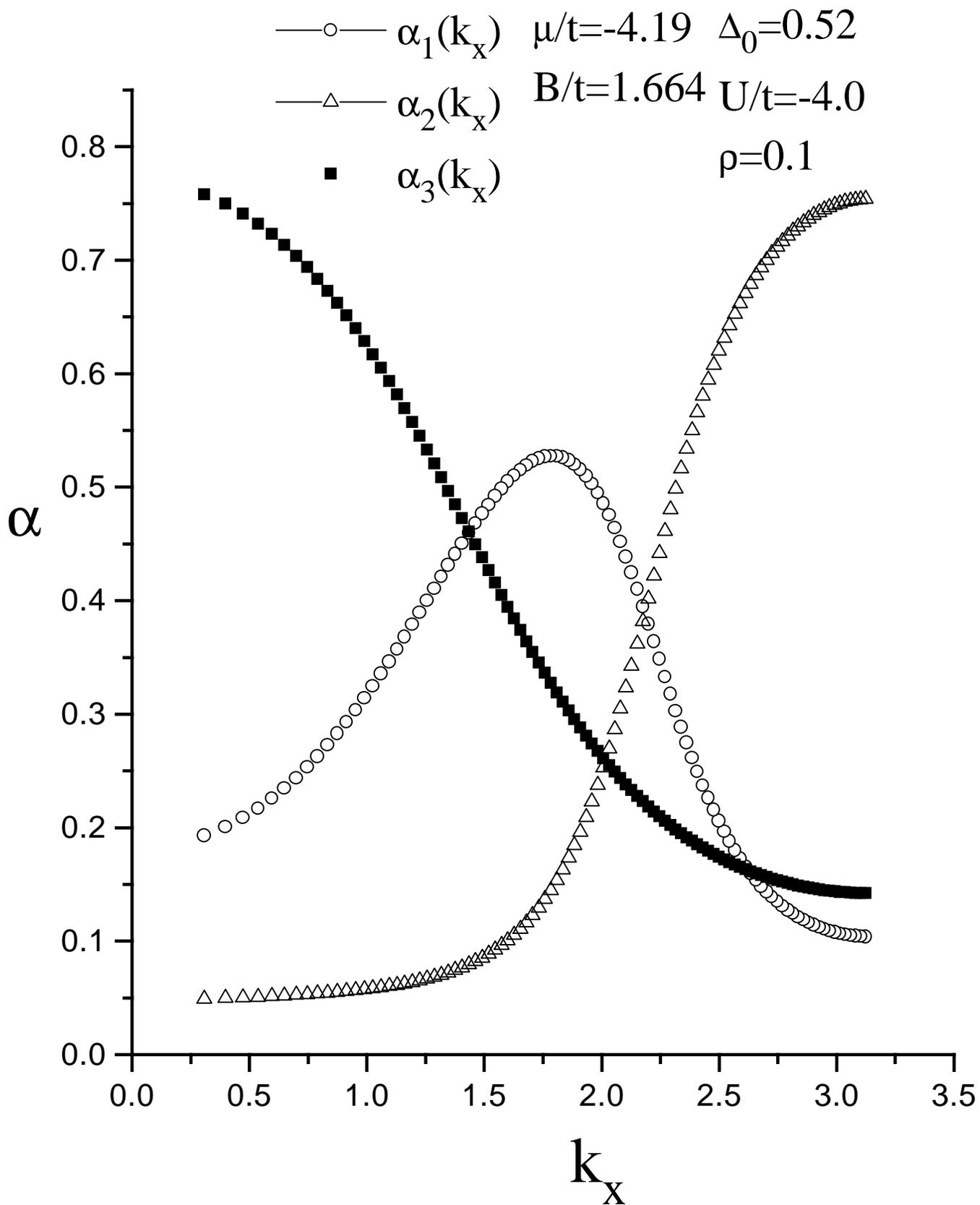
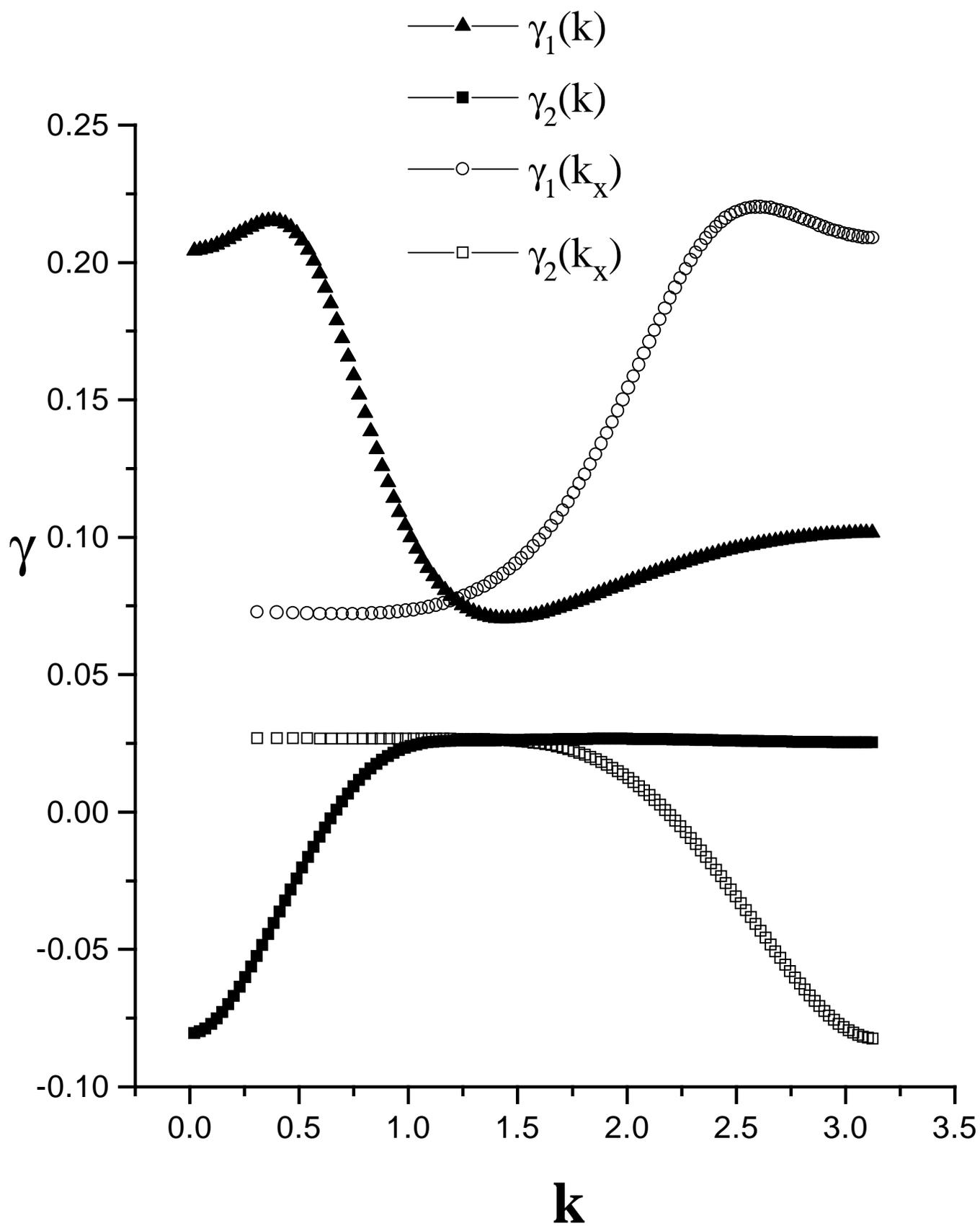
\end{document}